\documentclass[12pt]{emulateapj}
\usepackage[dvips]{color} 

\slugcomment{Not to appear in Nonlearned J., 45.}

\shorttitle{Inward motion of Markarian 421}
\shortauthors{Niinuma et al.}

\begin{document}


%\title{VLBI observation for the largest X-ray flare in the TeV blazar Mrk 421}
%\title{Possible detection of the inward and superluminal motion in the active galaxy Mrk 421}
\title{Possible Detection of Apparent Superluminal inward motion in Markarian 421 
after the Giant X-ray flare in February, 2010}


\author{K. Niinuma\altaffilmark{1}, M. Kino\altaffilmark{2}, H. Nagai\altaffilmark{3}, N. Isobe\altaffilmark{4}, K. E. Gabanyi\altaffilmark{5, 10}, K. Hada\altaffilmark{6}, S. Koyama\altaffilmark{7}, K. Asada\altaffilmark{8}, T. Oyama\altaffilmark{2}, and K. Fujisawa\altaffilmark{9}}

\altaffiltext{1}{Graduate School of Science and Engineering, Yamaguchi University, Yamaguchi 753-8511, Japan}
\altaffiltext{2}{Mizusawa VLBI Observatory, National Astronomical Observatory of Japan, Osawa, Mitaka, Tokyo 181-8588}
\altaffiltext{3}{ALMA-J Project, National Astronomical Observatory of Japan, Osawa, Mitaka, Tokyo 181-8588}
\altaffiltext{4}{Institute of Space and Astronautics, Japan Aerospace Exploration Agency, Yoshinodai, Chuo, Sagamihara 252-5210}
\altaffiltext{5}{Hungarian Academy of Sciences, Research Group for Physical Geodesy and Geodynamics, FOMI Satellite Geodetic Observatory
Budapest, 1592 Hungary}
\altaffiltext{6}{INAF  Istituto di Radioastronomia, via Gobetti 101, I-40129, Bologna, Italy}
\altaffiltext{7}{Department of Astronomy, The University of Tokyo, Hongo, Bunkyo-ku, Tokyo 113-8654}
\altaffiltext{8}{Academia Sinica Institute of Astronomy and Astrophysics, Taipei 10617, Taiwan}
\altaffiltext{9}{Research Institute for Time Studies, Yamaguchi University, Yamaguchi 753-8511, Japan}
\altaffiltext{10}{Konkoly Observatory, Research Centre for Astronomy and Earth Sciences, Hungarian Academy of Sciences, 1121, Budapest, Hungary}

\email{niinuma@yamaguchi-u.ac.jp}



\begin{abstract}
We report on the VLBI follow-up observations using the Japanese VLBI Network (JVN) array at 22 GHz for the largest X-ray flare of TeV blazar Mrk 421 that occurred in mid-February, 2010. The total of five epochs of observations were performed at intervals of about 20 days between March 7 and May 31, 2010. No new-born component associated with the flare was seen directly in the total intensity images obtained by our multi-epoch VLBI observations. However, one jet component located at $\sim1$ mas north-west from the core was able to be identified, and its proper motion can be measured as $-1.66\pm0.46$ mas yr$^{-1}$, which corresponds to an apparent velocity of $-3.48\pm0.97 c$. Here, this negative velocity indicates that the jet component was apparently moving toward the core. As the most plausible explanation, we discuss that the apparent negative velocity was possibly caused by the ejection of a new component, which could not be resolved with our observations. In this case, the obtained Doppler factor of the new component is around 10 to 20, which is consistent with the ones typically estimated by model fittings of spectral energy distribution for this source.
\end{abstract}


\keywords{BL Lacertae objects: individual (Mrk 421) --- techniques: interferometric}



%%%%%%%%%%%%%%%%%%%% INTRODUCTION %%%%%%%%%%%%%%%%%%%%%%

\section{Introduction}
Blazars are a sub-class of radio-loud active 
galactic nuclei (AGNs) with relativistic jets
pointing almost along the line of sight.
The spectral energy distributions (SEDs) of blazars are 
dominated by non-thermal, strongly Doppler boosted,
and variable radiation produced at the 
innermost part of the jets.
The SED and the multi-frequency correlations of blazars
have been intensively studied in the past through
dedicated multi-frequency campaign observations
\citep[eg.][and references therein]{abdo11}
%(e.g., Abdo et al. 2011 and reference therein).
%%%%

One of the most important pieces for understanding 
relativistic jet formation is to obtain the direct
image of a region radiating high energy emission.
However, it is difficult to obtain these images
 for the following reasons.
(1)
Many of the radio observations have been done by
single-dish telescopes \citep[eg.][]{raiteri09, villata09}
 and their spatial resolution is much larger than the relevant size.
(2)
Recent progress of VLBI observations indicate possible correlations between 
newborn VLBI components and high energy emissions by the MOJAVE 
(Monitoring Of Jets in Active galaxies with VLBI Experiment) 
project \citep[eg.][]{lister09a, kovalev09}. It is, however, difficult 
to confirm the correlation between a newborn VLBI component 
and X-ray and/or the very high energy (VHE) $\gamma$-ray flares 
when several flares occur during a single monitoring span which is 
typically a few months for each source.
Therefore, to confirm the fact mentioned above, 
to examine the behavior at the root of the jet by conducting a dense monitor 
using VLBI soon after ($\leq$ a few weeks) the giant X-ray and/or VHE 
$\gamma$-ray flare, can be a complementary approach.

Mrk 421 is one of the best sources for studying the most 
compact regions in blazars, because of its proximity ($z=0.031$, 131 Mpc).
With respect to the kinematics of the parsec-scale jets in Mrk 421,
 very slow and/or almost stationary components, which can be monitored by 
 long-term observations with VLBI, have been studied by \citet{piner10}. 
 However, no dense monitoring has yet been conducted to study
  the behavior of near the core of Mrk 421 right after the giant flares (see, \S \ref{flares}).

Throughout this paper, we adopt the cosmological parameters
of $H_0=71$ km s$^{-1}$ Mpc$^{-1}$, $\Omega_m=0.27$, and 
$\Omega_{\Lambda}=0.73$ \citep{komatsu09}. Using these parameters, 
1 mas corresponds to a linear distance of 0.64 pc, and a proper
 motion of 1 mas yr$^{-1}$ corresponds to an apparent velocity 
of $2.1 c$ at the distance of Mrk 421.

The organization of this paper is as follows: In \S \ref{flares}, we summarize past giant flares of Mrk 421, and their follow-up observations using VLBI. In \S\ref{obs_reduc}, \S\ref{modelfit}, and \S\ref{result}, we describe our observations, data analysis, and the results obtained from model-fitting the data. In \S\ref{mojave}, we compare our result with the one derived from the MOJAVE program to identify the position of the jet component, and in \S\ref{discussion}, we discuss the astrophysical implications derived from our result.

%%%%%%%%%%%%%%%%%%%%%%%% Review of Flare %%%%%%%%%%%%%%%%%%%%%%%%
\section{Review of giant VHE $\gamma$-ray and X-ray flares}\label{flares}
As mentioned in \S 1, systematic VLBI monitor within $\sim100$ days for investigating 
the behavior of near the core after the occurrence of a giant flare have not done yet. 
In this section, we briefly review the previous giant flares of Mrk 421 at VHE and/or X-ray band,
 and the follow-up observations with VLBI for the flares.
 
\subsection{Previous flares}
%\kino{To} discuss the difference between the previous studies and our paper, we summarize the large \kino{VHE $\gamma$-ray / X-ray} flares which were detected in Mrk 421, and the VLBI follow-up observations carried out after the occurrence of such flares.
%
Since the first discovery of VHE $\gamma$-ray emission from Mrk 421 was reported by \citet{punch92}, large X-ray or VHE $\gamma$-ray flares of Mrk 421 have occasionally been reported. \citet{gaidos96} succeeded in detecting the VHE $\gamma$-ray flares from Mrk 421 within a few years after the discovery of VHE emission of Mrk 421. An extremely high state of Mrk 421 at VHE band, which lasted from 2001 January to March was reported by \citet{krennrich02}. Recently, multi-band  observations have also been performed for such large flares of Mrk 421 \citep{blazejowski05,lichti08,donnarumma09}

In Table \ref{tbl1}, we listed the large flares of Mrk 421, and their VLBI follow-up observations. (Here, we excluded the information of geodetic observations). The epoch, peak fluxes at X-ray/VHE band, and references of each flare are indicated in column 1-4, and the information about VLBI observations for these flares is shown in column 5-7, if the observations were done within three months after the flare occurred. Columns 5 and 6 show their first epoch (as the earliest one), and the averaged interval of each epoch in the case that multi-epoch observations were done, respectively. For example, although \citet{piner05} carried out the multi-epoch observations for the exceptionally strong TeV gamma-ray flare in 2001 with VLBA, their first epoch was done about five months after the flare and the intervals of each observation were about three months.

\subsection{The flares detected by MAXI}
The total of four strong flares ($\geq$ 100 mCrab) of Mrk 421 were detected with the energy ranging from 2 keV to 10 keV between 2009 November 1 and 2010 February 16 \citep{isobe10} by the Monitor of All-sky X-ray Image \citep[MAXI; ][]{matsuoka09}, on the Exposed Facility of the Japanese Experiment Module "Kibo" attached to the International Space Station (ISS). The second and fourth ones were also clearly detected with the energy range of Swift/BAT (15 keV - 50 keV). In the fourth one, the flux reached $164\pm17$ mCrab on a 6-hour average which is the largest one at X-ray band among those ever reported. Especially for the largest X-ray flare, the detection of very bright VHE $\gamma$-ray emission ($\sim10$ Crab) was also reported \citep{ong10}.%Recently, \citet{shukla12} reported on the result of multi-band study of Mrk 421 in this very high state.
%%
%%%%%%%%%%%%%%%%%%%%%%%% Observations %%%%%%%%%%%%%%%%%%%%%%%%
\section{Observations and data reduction}\label{obs_reduc}
Anticipating the detection of a newly born component associated with the largest X-ray flare, we started follow-up observations with the Japanese VLBI network \citep[JVN:][]{doi06}\footnote{\url{http://www.astro.sci.yamaguchi-u.ac.jp/$\sim$kenta/jvnhp/eng/index\_e.html}} within a month after 2010 February flare. JVN includes VLBI Exploration of Radio Astrometry \citep[VERA:][]{kobayashi03}, and is spread across the Japanese islands. Our first JVN observation of Mrk 421 took place on 2010 March 7. It was followed by another 4 epochs distributed by regular intervals of $\sim20$ days. The last observation was conducted on 2010 May 31. All these observations were performed at 22 GHz. Unfortunately the data we could analyze are only from VERA antennas because the other antennas mainly suffered from recording trouble.

The data from 4 telescopes ("VERA Mizusawa", "VERA Iriki", "VERA Ogasawara", and "VERA Ishigaki-jima": Baseline ranging over $1000-2300$ km) that were recorded with a bandwidth of 256 MHz using 2-bit samples and left-circular polarization, were used for data reduction. The on-source time of Mrk 421 in each epoch is approximately 5.5-hr, and the range of elevation at which we carried out observations for Mrk 421 were $35\degr \leq El \leq 88\degr$ until the fourth epoch, and $10\degr \leq El \leq 82\degr$ at the fifth epoch. The elevation dependence of the aperture efficiency of each VERA antenna is almost flat at $El \geq 25\degr$\footnote{\url{http://veraserver.mtk.nao.ac.jp/restricted/CFP2011/status11.pdf}}. The synthesized beam size in our observations is typically a major axis of 1.2 mas and a minor axis of 0.8 with a position angle of $-50\degr$ measured from north through east. The correlation was carried out on the Mitaka FX correlator \citep{chikada91}.

In the last epoch May 31, fringes to the "VERA Mizusawa" antenna could not be detected due to system troubles. This means the maximum baseline length was lacking. Therefore we only had data from three antennas in this epoch. Since the synthesized beam size is 1.5 times larger than in the other four epochs, we decided to exclude this observation from further analysis. Also, at the second epoch Apr 1, Mrk 421 was not detected in the more than half of an observing time at the baseline including VERA Iriki station because of bad weather.

For the correlated data, we performed the standard VLBI calibration and fringe-fitting procedure using the NRAO Astronomical Image Processing System (AIPS) software package. The flux calibration done by the AIPS task APCAL typically achieved an accuracy of $10\%$, for VERA at 22 GHz \citep{nagai10}. The calibrated visibilities were then exported to carry out an imaging procedure using the Caltech Difmap package. After the time averaging of visibility data in 30-second bins, we carried out the iteration of CLEAN and self-calibration procedures in Difmap. During this iteration process, the solution interval of an amplitude self-calibration was shortened from intervals as long as the whole observational time down to 15 minutes.

%%%%%%%%%%%%%%%%%%%% MODEL FIT %%%%%%%%%%%%%%%%%%%%%%
\section{Modelfit}\label{modelfit}
\begin{figure*}[hbtp]
\centering
\includegraphics[scale=0.85]{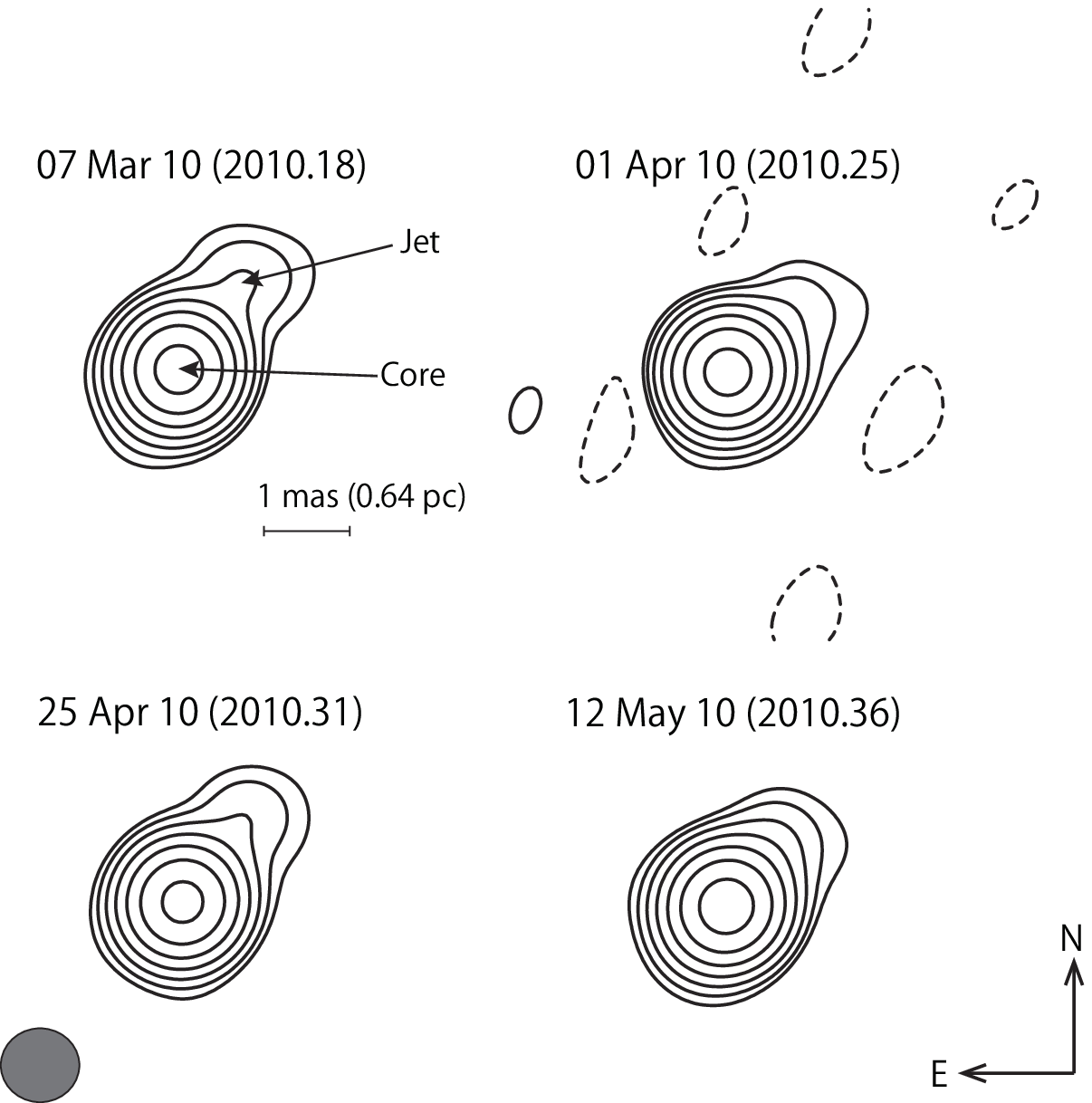}
\caption{VLBI images of Mrk 421 with natural weighting at 22 GHz. Images are arranged from top left to bottom right in order of epoch. The epoch is indicated on the top of each panel as "DD MMM YY (YYYY.YY)". Also, the scale of 1 mas for our images is indicated on the panel of 1st epoch, two arrows on the image of 1st epoch show the core and the jet feature. These images are convolved with a circular beam size (full width at half maximum: FWHM) of 0.8 mas (shown in the bottom-left corner), which corresponds to the typical minor axis of synthesized beam in our observations. Contours begin at 3 mJy beam$^{-1}$ and increase in -1, and $2^{n}$ steps. The peak flux and the image rms ($F_{peak}$ mJy beam$^{-1}$, $F_{rms}$ mJy beam$^{-1}$) at each epoch are (257.6, 0.6), (253.8, 1.5), (236.8, 0.5), (278.9, 0.8), respectively.\label{fig:fig1}}
\end{figure*}

The total intensity images of Mrk 421 are shown in Figure \ref{fig:fig1}. These images were convolved with a circular beam with an axis size (full width at half maximum) of 0.8 mas (shown in the bottom left corner of the figure), which corresponds to the typical minor axis of a synthesized beam in our observations. Our observations confirmed a north-west oriented core-jet structure of Mrk 421. The direction of the outflow is consistent with those reported in other studies \citep{lico12, piner99, piner05, piner10}.

Following \citet{piner10}, we have carried out the elliptical and the circular Gaussian model-fitting to the self-calibrated visibility data in the ($u-v$)-plane using the task {\it modelfit} in the Difmap at all epochs. In such a way as to avoid the unrealistic changes in Gaussian sizes between adjacent epochs, we found that the elliptical Gaussian model for the core and the circular one for the jet component named JC1 is an appropriate combination. Model parameters are given in Table \ref{tbl2}. The $1\sigma$ errors or $1\sigma$ upper limits of each parameter were derived from the model-fit procedure. The parameter we focused on is varied around the best-fit value by confirming a change in the $\chi^2$ value. The position angle of the core components is around $-30\degr$, which coincides with that between the core and JC1 (about $-40\degr$ except for the 2nd epoch). From our model-fit, the model components for the core were resolved toward the jet axis at all epochs. Regarding the direction perpendicular to the position angle of the core component, one is resolved at only the 3rd epoch, and others are unresolved. When we fixed the axial ratios of the core components as the value obtained at the 3rd epoch, a slight difference between the major axis of the core component and the jet axis generates a negligibly small fitting error of 0.02 mas at maximum. Therefore JC1 is sufficiently distinguished from the core at all epochs, because of the size of its major axis.

%%%%%%%%%%%%%%%%%%%% RESULTS %%%%%%%%%%%%%%%%%%%%%%
\section{Results}\label{result}
\begin{figure}[hbtp]
\centering
\includegraphics[scale=0.6]{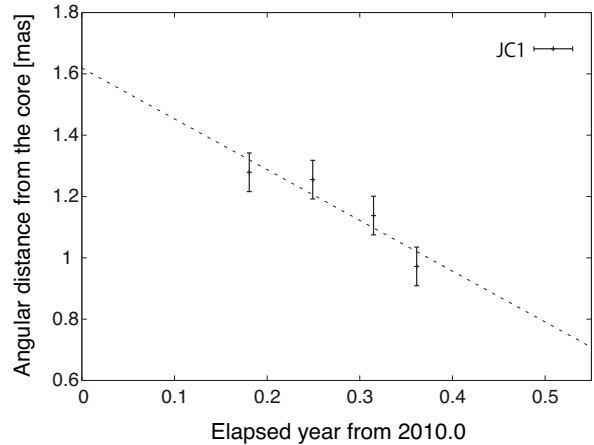}
\caption{JC1 positions and weighted linear fit to JC1 positions ({\it dashed line}). The position uncertainties of each data point are uniformly rescaled so that the reduced $\chi^2$ to be unity. As the result of least square fit, we obtained the proper motion of $-1.66\pm0.46$ mas yr$^{-1}$.\label{fig:fig2}}
\end{figure}

The position of JC1 relative to the core is plotted as a function of time in Figure \ref{fig:fig2}. As the result of the least square fit to the displacements of the position of JC1 relative to the core during 66 days (from Mar 7 to May 12) under the assumption of linear motion of JC1, we obtained the proper motion of $\mu_{1a}=-1.66\pm0.46$ mas yr$^{-1}$ which corresponds to the apparent velocity of $v_{1a}=-3.48\pm0.97 c$ at the distance of Mrk 421. This negative velocity implies apparent inward motion of JC1. To measure the apparent velocity of JC1, although we have to estimate not only thermal, but systematic errors as the uncertainty on the position of JC1 relative to the core at each epoch, it is quite difficult to estimate the latter quantitatively due to its non-linearity. Therefore, following \citet{homan01, piner10}, we set the positional uncertainty on the data points that were uniformly rescaled so that the reduced $\chi^2$ is unity. Also, as seen in Figure \ref{fig:fig2}, if we exclude the last epoch from the fit, the result obtained would only be half of the value mentioned above. Here, we applied the first order fit to the motion of JC1. Based on our results, the significance of the detected superluminal motion of JC1 in Mrk 421 is $3.57 \sigma$.

On the other hand, though the 10\% flux errors mentioned in \S \ref{obs_reduc} are slightly large relative to the core flux change, which can be seen in Table \ref{tbl2}, it appears that the core flux in the last epoch is higher than in the previous epochs. Additionally, the other flux data, which were compiled by F-GAMMA program\footnote{\url{http://www.mpifr-bonn.mpg.de/div/vlbi/fgamma/results.html}}\citep{angelakis10}, also shows almost the same trend of increasing flux as our result. Although the scale of flux variability was not large, there is a possibility that a delayed increase in radio flux was seen. When the flare occurs in the blazars, it has often been reported in the radio observations at low frequency because of the opacity effect \citep[e.g.,][]{orienti11}. These support the possibility of structural changes occurring in the core region, such as the emergence of a new component. The details of the flux density change at radio band after the X-ray flare of 2010 February, will be discussed in a forthcoming paper. 

%%%%%%%%%%%% Identification of JC1 %%%%%%%%%%%%%%
\section{Identification of JC1}\label{mojave}

To identify the correspondence between the jet component JC1 and the one detected by other observations, we analyzed the data observed with VLBA at 15 GHz \citep[MOJAVE program:][]{lister09c}. We obtained the calibrated visibility data at a total of five epochs (2009 Dec 17, 2010 Feb 11, Jul 12, Oct 15, and Nov 29) from the MOJAVE web site, and carried out the standard analysis procedure. In Figure \ref{fig:fig4}, we show a CLEANed image within 4 mas from the core at the epoch of 2010 February 12th. Also, model fit parameters are shown in Table \ref{tbl3}. Here, we carried out the model fit in the same way as mentioned above. As the result, we identified four jet components in this scale by comparing the angular distance from the core, the position angle, the size, and the flux density of each model throughout five epochs, and named them MC1, MC2, MC3, and MC4 from outermost to innermost component. In addition, as temporary components, MC3a and MC4a are seen in the third and fifth epoch, respectively.

\begin{figure*}[hbtp]
\centering
\includegraphics[scale=0.8]{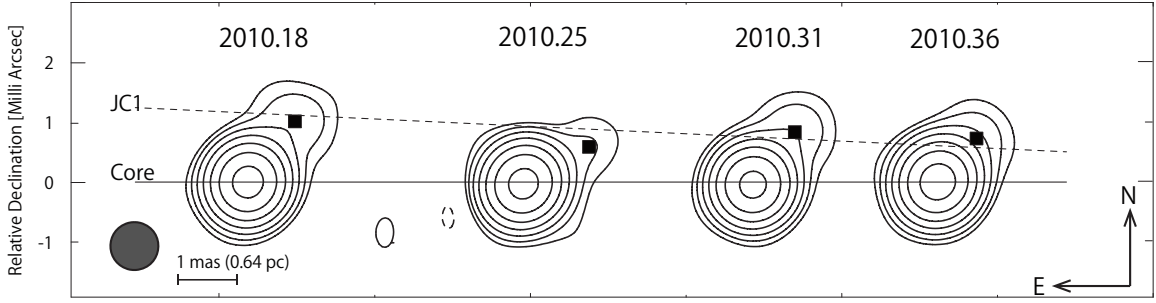}
\caption{VLBI images of Mrk 421 produced by restoring model-fit components. The epoch and the scale of 1 mas for our images are indicated on the top of each image as year "YYYY.YY", and in the bottom of the panel of the 1st epoch, respectively. The interval of each image is proportional to the date when each observation was made. The filled squares represent the position of JC1 relative to the core, and the solid and the dashed lines show the trajectory of  the core and of JC1, respectively. The decrease in the separation between the core and JC1 is shown by these two lines. The beam size of this figure is the same as Figure \ref{fig:fig1}, and is shown in the bottom-left corner of the panel of the 1st epoch. Contours begin at 3 mJy beam$^{-1}$ and increase in -1, and $2^{n}$ steps.\label{fig:fig3}}
\end{figure*}

\begin{figure}[hbtp]
\centering
\includegraphics[scale=0.6]{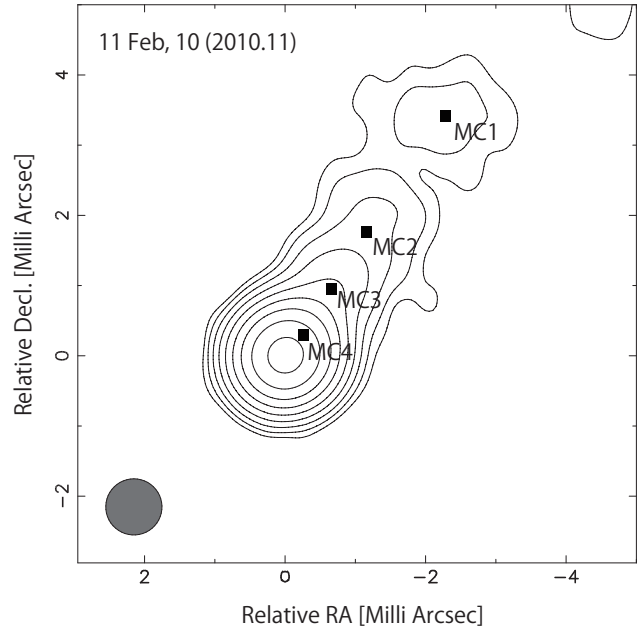}
\caption{VLBA image of Mrk 421 with natural weighting at 15 GHz (MOJAVE program).  The epoch is indicated on the top of each panel as "DD MMM, YY (YYYY.YY)", and the filled squares represent the position of each component relative to the core. The beam size of this figure is the same as Figure \ref{fig:fig1}, and is shown in the bottom-left corner of the panel of the 1st epoch. Contours begin at 0.9 mJy beam$^{-1}$ and increase in $2^{n}$ steps.\label{fig:fig4}}
\end{figure}

\begin{figure}[hbtp]
\centering
\includegraphics[scale=0.6]{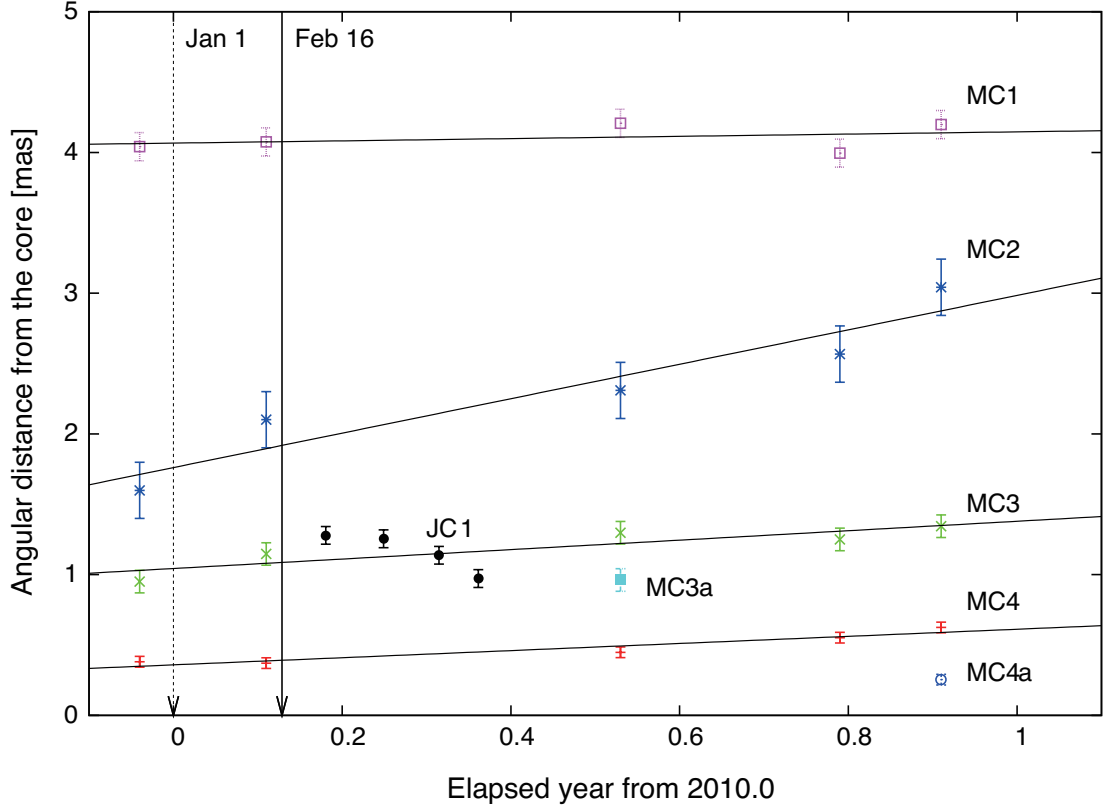}
\caption{Angular distances of each component detected with our observations and MOJAVE, relative to the core. The dashed and the solid arrows indicate the date when the large X-ray flares occurred. Also, each solid line shows the results of the linear-fit to each component without JC1.\label{fig:fig5}}
\end{figure}

The angular distance from the core for all model components at each epoch derived from the MOJAVE data, and the one detected by our observations (JC1) are shown in Figure \ref{fig:fig5}. From the model fit parameters, we can identify the JC1 as MC3. MC1, MC2, and MC4 cannot be detected in the JVN images (e.g., Figure \ref{fig:fig3}). The reason MC1 and MC2 were not detected is that the image sensitivity of our observations was insufficient to detect both components which are relatively faint. And MC4 cannot be resolved with the angular resolution of our observations either, because of its closeness to the core.

Each solid line in Figure \ref{fig:fig5} indicates the least square fit to the motion of each component using first order fit. These fitting were done without JC1 data, and each error bar was calculated in the same way as in Figure \ref{fig:fig2}. For MC3a and MC4a, each positional error is set the same value as MC3 and MC4 respectively, because each component was not detected at other epochs. In this fit, the components showed the following proper motion: for MC1 $0.08\pm0.12$ mas yr$^{-1}$, for MC2 $1.22\pm0.24$ mas yr$^{-1}$, for MC3 $0.34\pm0.10$ mas yr$^{-1}$, and for MC4 $0.25\pm0.05$ mas yr$^{-1}$, which correspond to $0.17\pm0.25 c$, $2.56\pm0.50 c$, $0.71\pm0.21 c$, and $0.53\pm0.11 c$, respectively. MC1, MC3, and MC4 showed the sub-luminal speeds, which are consistent with the results shown in previous works \citep[e.g.,][]{piner10}. In these five epochs, MC2 showed super-luminal motion. Recently, \citet{lico12} also performed VLBA observations of Mrk 421 every month in 2011. In their 15 GHz observations, four jet components are also detected, and their angular distance from the core was consistent with MC1, MC2, MC3, and MC4.

%%%%%%%%%%%% DISCUSSION %%%%%%%%%%%%%%
\section{Discussion}\label{discussion}
To examine the behaviour inside the sub-pc region of Mrk 421, we conducted multi epoch VLBI observations anticipating a super-luminal component that is ejected from the core region after a large flare. %%%
Instead of detecting a newly emerging component on the VLBI images, we detected the apparent super-luminal inward motion of the jet component JC1.
%%
%\kino{Although no newly born component can be seen in our VLBI images, the motion of the jet component JC1 showed systematic %inward motion at a possibly super-luminal speed rather than the positional fluctuation of JC1 in short span \citep{edwards02}.}
%%
%%%
Such motions of jet components have been reported by previous MOJAVE project studies in which \citet{lister09b} discussed the detection probability of super-luminal inward motions among all observations to be approximately 2\%. We consider our detection of the inward motion of JC1 in Mrk 421 is similar to the one rarely seen in the MOJAVE samples.

%As discussed in \citet{edwards02}, the possibility that the bobbling of jet component close to the core in a short time span was shown as an inward motion of JC1 is also considerable. To confirm it, however, we believe that the long term and densely sampling (within a couple of weeks) is needed. Additionally, it seems that JC1 moved in closer to the core systematically through the first to the last epochs. 
There is also a possibility that the positional fluctuations of the jet component close to the core showed its unusual motion in short span\citep{edwards02}, however, the motion of JC1 showed systematic inward motion at a super-luminal speed rather than its positional fluctuations. Therefore, here we discuss several possible ideas suggested in \citet{kellermann04} for explaining an apparent super-luminal inward motion of JC1. 
Among them, it is clear that 
the free-free absorption effect of disk, torus, and ambient 
cold plasma \citep[the case discussed in][]{kameno01} seems 
less likely.
%%%
Another possibility they indicated is a highly curved jet. 
Although such extreme bending has been reported in
 a small number of sources \citep[see, e.g.,][]{savolainen06},
 previous higher resolution studies for Mrk 421 do indeed not
  show any evidence for extreme bending within a parsec of the jet.
%But it seems artificial to impose a jet with such extreme bending
% within a parsec of the jet.
%%
And, the possibility of the bobbling of the core component with a short time span
 is also considerable. However, it seems that JC1 moved in closer to the core systematically
 through the first to the last epochs.
Therefore, below, we focus on the following scenario proposed by Kellerman:
a newly born component (JC0) on 16th Feb 2010, when the largest X-ray flare occurred, 
was in the core but it was not resolved by our beam and it made the centroid of the core
 shift toward the jet motion.
%%%
 Also, as seen In Figure \ref{fig:fig5}, the motions of JC1 and MC3 seem discontinuous 
 between the third and the forth epoch of JVN observations (Apr 25 and May 12, respectively), 
 and the third epoch of MOJAVE observation (Jul 12). 
%%% 
 It is, however, possible to explain this motion as JC1 (or MC3) got back on the track confirmed by 
 previous studies, because the newborn component associated with the giant flare, which 
 shifted the centroid of the core faded within approximately 0.2 years.
%%%
Therefore, basically, we have considered that the motion and the position of JC1 detected 
by our observations were consistent with previous studies.

\begin{figure}[hbtp]
\centering
\includegraphics[scale=0.6]{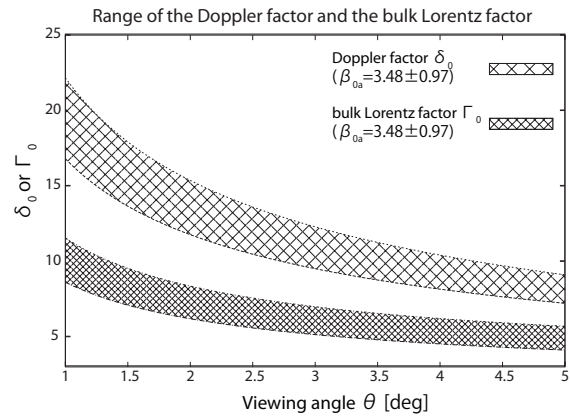}
\caption{$\delta_0$ and $\Gamma_0$ derived from the apparent velocity of the newborn VLBI component JC0 with the viewing angle ranging $1\degr\leq\theta\leq5\degr$. At the viewing angle of $\theta=1\degr$, $\delta_0$ and $\Gamma_0$ exceed 16.8, 8.6, respectively.\label{fig:fig6}}
\end{figure}

Lastly, we discuss the Doppler factor and the Lorentz factor of JC0 under the assumption mentioned above. The apparent velocity of $v_{1a}=-3.48\pm0.97 c$ derived from the linear fit to the positions of JC1 relative to the core, is the relative velocity. Therefore assuming that JC1 did not move or its apparent velocity is sufficiently slow, we can consider that the JC0 moved toward JC1 with the apparent velocity of $v_{0a}\sim3.48\pm0.97 c$. In Figure \ref{fig:fig6}, we show the bulk Lorentz factor ($\Gamma_{0}$) and the Doppler factor ($\delta_{0}$) of JC0 calculated by setting a range of $2.51c\leq v_{0a}\leq 4.45c$, and $1\degr\leq\theta\leq5\degr$, respectively. Here we use the equations of $\Gamma_{0} = 1/\sqrt{1-\beta_{0}^2}, \delta_{0}=1/[\Gamma_{0}(1-\beta_{0}\cos\theta)]$, (where, $\beta_0=v_0/c=\beta_{0a}/[\sqrt{1+\beta_{0a}^2}\sin(\theta+\phi)], \beta_{0a}=v_{0a}/c, \phi=\cos^{-1}(1/\sqrt{1+\beta_{0a}^2})$). As the result of these calculations, $\delta_{0}=11.03^{+1.26}_{-1.52}$, and $\Gamma_{0}=6.11^{+0.88}_{-0.97}$ are obtained for the median value of $\theta=3\degr$. 

We consider that these values are reasonable, because our result does not require extreme values for the jet parameters and the Doppler factor $\delta_0$ is almost consistent with the typical value of $\sim10$ derived from the broad band (optical to TeV $\gamma$-ray) SED fit \citep[eg.][]{kino02, blazejowski05, donnarumma09, abdo11}.

%%%%%%%%%%%%%%%%%%%% CONCLUSION %%%%%%%%%%%%%%%%%%%%%
\section{Conclusion}\label{conclusion}
For giant X-ray flare of Mrk 421 occurred on 2010 February 16, we carried out the multi-epoch VLBI observations at intervals of $\sim 20$ days using JVN array soon after the giant flare. Although no new-born component can be seen directly in the total intensity images obtained by our quick follow-up observations, the jet component JC1 located at $\sim 1$ mas north-west from the core showed its apparent inward motion at an apparent speed of $3.48\pm0.97 c$. As the most plausible explanation, we discuss that the inward motion of JC1 at a super-luminal speed was possibly caused by the ejection of a new component JC0, which could not be resolved with our observations. In this case, the obtained Doppler factor of JC0 is around 10 to 20, which is consistent with the ones typically estimated by model fittings of spectral energy distribution for this source.

%The 2010 February flare, for which we conducted follow-up observations using VLBI, exhibited the largest X-ray flux \citep{isobe10}. This flare was presumably not so special one. If multi-epoch VLBI observations are performed for the giant flare right after it occurs at a short interval (within a few weeks), we have believe that there is a possibility that the newly born component can be detected, or the inward motion as we detected can be seen. We have also considered that the reasons why \citet{piner05} found no newly born component even though they carried out multi-epoch VLBI observations for the extreme large flare of Mrk421, is due to the timing of the start of their observations and the intervals of them that were not sufficiently early and sufficiently dense to catch a \kino{newborn} component associated with it.

To confirm our result, we must carry out a quick follow-up or sufficient densely monitoring for such a large flare with a phase-referencing VLBI technique. By doing this, there is a possibility of identifying the actual motion of all components including the core region after the large flare. Also, a polarization observation such as \citet{piner05}, or \citet{marscher08} is also an effective means of identifying whether a VLBI component is produced by the flare, and to investigate its motion right after the flare occurs.

%%%%%%%%%%%%%%%%%%%% Acknowledgement %%%%%%%%%%%%%%%%%%%%%
\acknowledgments
We are grateful to the anonymous referee whose suggestions improved our paper substantially. We are grateful to all JVN members for their assistance in observations. We also thank K. Akiyama for very useful comments for the data reduction. The JVN project is led by the National Astronomical Observatory of Japan, which is a branch of the National Institutes of Natural Sciences, Hokkaido University, The University of Tsukuba, Ibaraki University, Gifu University, Osaka Prefecture University, Yamaguchi University, and Kagoshima University, in cooperation with the Geographical Survey Institute, the Japan Aerospace Exploration Agency, and the National Institute of Information and Communications Technology. This research has made use of data from the MOJAVE data base that is maintained by the MOJAVE team \citep{lister09c}. KN acknowledges financial support from Grant-in-Aid for Young Scientists (B) (No. 22740130) from Japan Society for the Promotion of Science (JSPS). This work is partially supported by Grant-in-Aid for Scientific Research, KAKENHI 24540240 (MK) from Japan Society for the Promotion of Science (JSPS).. K. E. G. was supported by the Hungarian Scientific Research Fund (OTKA, grant no. K72515).

\begin{table}[htbp]
\caption{Large X-ray and/or VHE $\gamma$-ray flares of Mrk 421 and VLBI follow-up\label{tbl1}}
\centering

\begin{tabular}{lcccclcc} \hline
\hline
\multicolumn{4}{c}{X-ray or VHE $\gamma$-ray flare} &       & \multicolumn{3}{c}{Astrophysical VLBI observations} \\
\cline{1-4} \cline{6-8}
Epoch & $F_{{\rm peak, VHE}}^{{\rm a}}$ & $F_{{\rm peak, X}}^{{\rm b}}$ & Ref. &   & First epoch & Intervals & Ref. \\
             & [Crab] & [mCrab] &  &  &  & & \\ \hline
%%1994 May         & $<3$ & & 1 &   & -- & -- & -- \\
1996 May         & $\sim10$ & $<30$ & 1, 2 &   & 1996 May & -- & 12 \\
2001 Jan-Mar & $\sim13$ & -- & 3 &   & 2001 Mar & 3 - 5 months & 12, 13 \\
2004 Mar-Apr & $\sim3$ & $\sim135$ & 4 &   & -- & -- & -- \\
2006 Jun         & -- & $\sim92$ & 5, 6 &   & 2006 Jun & -- & 12 \\
2008 Mar         & -- & $\sim110$ & 7 &   & 2008 May & $\sim2$ months & 12 \\
2008 Jun         & -- & $\sim125$ & 8 &   & 2008 Jun & $\sim2$ months & 12 \\
2009 Nov        & -- & $\sim96$ & 9 &   & 2009 Dec & 3 months & 12 \\
2010 Jan & -- & 120$\pm$10 &  10 &   & 2010 Jan & 3 months & 12 \\
2010 Feb & $>10$ & 164$\pm$17 &  10, 11 &   & 2010 Mar & $\sim20$ days & 14 \\
 % &  &   & -- & -- & -- \\ \hline
\hline
  
\end{tabular}
\tablenotetext{a}{The peak flux at VHE $\gamma$-ray band}
\tablenotetext{b}{The peak flux at X-ray band}
\tablenotetext{}{Reference -- 1: \citet{gaidos96}, 2:\citet{schubnell97}, 3: \citet{krennrich02}, 4: \citet{blazejowski05}, 5: \citet{lichti08}, 6: \citet{lichti06}, 7: \citet{krimm08}, 8: \citet{donnarumma09}, 9: \citet{dammando09}, 10: \citet{isobe10}, 11: \citet{ong10}, 12: MOJAVE program (\url{http://www.physics.purdue.edu/astro/MOJAVE/index.html}), 13: \citet{piner05}, 14: This study}
\end{table}

%% If you use the table environment, please indicate horizontal rules using
%% Do not put multiple tabular environments within a single table.
%% The optional \label should appear inside the \caption command.

\begin{table*}[htbp]
\caption{Model fit parameters at each epoch - JVN 22 GHz.\label{tbl2}}
\centering

\begin{tabular}{ccccccccc} \hline
\hline
Epoch & Component & \multicolumn{ 4}{c}{Model parameter} &       & \multicolumn{ 2}{c}{Distance from the core} \\
\cline{3-6} \cline{8-9}
      &       & Maj   & Axial Ratio & PA    & Flux Density &       & R     & $\theta$ \\
      &       & [mas] &       & [deg] & [mJy] &       & [mas] & [deg] \\
\hline
2010 Mar 07 & Core  & 0.23$\pm0.02$ & $<0.27$ & -24.0$\pm4.5$ & 263.7 &       & -     & - \\
                & JC1    & 0.65$\pm0.01$ & 1.0     & -     & 17.9 &       & $1.28\pm0.05$ & -37.6$\pm1.5$ \\
      &       &       &       &       &       &       &       &  \\
2010 Apr 01 & Core  & 0.29$^{+0.03}_{-0.06}$ & $<0.35$     & -36.7$^{+11.5}_{-9.5}$ & 261.5 &       & -     & - \\
                & JC1    & 0.51$^{+0.35}_{-0.51}$ & 1.0     & -     & 9.1 &       & $1.26\pm0.17$ & -60.4$\pm7.5$ \\
      &       &       &       &       &       &       &       &  \\
2010 Apr 25 & Core  & 0.18$\pm0.02$ &  0.37$^{+0.15}_{-0.30}$ & -34.8$^{+10.6}_{-10.3}$ & 242.2 &       & -     & - \\
                & JC1    & 0.60$\pm0.06$ & 1.0     & -     & 16.8 &       & $1.14\pm0.04$ & -38.3$\pm1.4$ \\
      &       &       &       &       &       &       &       &  \\
2010 May 12 & Core  & 0.26$\pm0.02$ & $<0.27$     & -34.0$\pm4.7$ & 288.5 &       & -     & - \\
                 & JC1    & 0.57$\pm0.11$   & 1.0   & -     & 14.3 &       & $0.97\pm0.06$ & -41.5$\pm2.7$ \\ \hline
\end{tabular}
\end{table*}

\begin{table*}[htbp]
\caption{Model fit parameters at each epoch - VLBA 15 GHz (MOJAVE).\label{tbl3}}
\centering

\begin{tabular}{ccccccccc} \hline
\hline
Epoch & Component & \multicolumn{ 4}{c}{Model parameter} &       & \multicolumn{ 2}{c}{Distance from the core} \\
\cline{3-6} \cline{8-9}
      &       & Maj   & Axial Ratio & PA    & Flux Density &       & R     & $\theta$ \\
      &       & [mas] &       & [deg] & [mJy] &       & [mas] & [deg] \\
\hline
2009 Dec 17 & Core  & 0.10 & 0.439 & -23.6 & 225.8 &       & -     & - \\
      & MC4    & 0.24 & 1.0     & -     & 23.3 &       & 0.38 & -31.7 \\
      & MC3    & 0.37 & 1.0     & -     & 16.0 &       & 0.95 & -35.1 \\
      & MC2    & 0.63 & 1.0     & -     &  8.3 &       & 1.60 & -32.9 \\
      & MC1    & 1.27 & 1.0     & -     &  9.5 &       & 4.04 & -32.5 \\
      &       &       &       &       &       &       &       &  \\
2010 Feb 11 & Core  & 0.10 & 0.00 & -73.1 & 279.2 &       & -     & - \\
      & MC4    & 0.24 & 1.0     & -     & 40.9 &       & 0.39 & -41.2 \\
      & MC3    & 0.42 & 1.0     & -     & 14.9 &       & 1.17 & -35.6 \\
      & MC2    & 0.97 & 1.0     & -     & 10.5 &       & 2.12 & -33.5 \\
      & MC1    & 1.16 & 1.0     & -     &  8.0 &       & 4.09 & -33.7 \\
      &       &       &       &       &       &       &       &  \\
2010 Jul 12 & Core  & 0.11 & 0.73 & -52.5 & 250.9 &       & -     & - \\
      & MC4    & 0.38 & 1.0     & -     & 35.4 &       & 0.46 & -42.4 \\
      & MC3a   & 0.16 & 1.0     & -     & 11.3 &       & 0.96 & -50.0 \\
      & MC3    & 0.52 & 1.0     & -     & 23.0 &       & 1.31 & -37.6 \\
      & MC2    & 0.46 & 1.0     & -     &  3.6 &       & 2.32 & -29.4 \\
      & MC1    & 1.40 & 1.0     & -     &  9.9 &       & 4.22 & -31.6 \\
      &       &       &       &       &       &       &       &  \\
2010 Oct 15 & Core  & 0.14 & 0.54 & -30.5 & 265.2 &       & -     & - \\
      & MC4    & 0.29 & 1.0     & -     & 22.7 &       & 0.56 & -39.1 \\
      & MC3    & 0.46 & 1.0     & -     & 17.6 &       & 1.26 & -37.6 \\
      & MC2    & 1.15 & 1.0     & -     &  8.9 &       & 2.58 & -37.7 \\
      & MC1    & 0.58 & 1.0     & -     &  4.6 &       & 4.00 & -29.6 \\
      &       &       &       &       &       &       &       &  \\
2010 Nov 29 & Core  & 0.11 & 0.73 & -51.9 & 248.4 &       & -     & - \\
      & MC4a  & 0.27 & 1.0     & -     & 56.6 &       & 0.25 & -26.3 \\
      & MC4    & 0.23 & 1.0     & -     & 17.3 &       & 0.67 & -39.5 \\
      & MC3    & 0.59 & 1.0     & -     & 16.3 &       & 1.38 & -40.2 \\
      & MC2    & 1.04 & 1.0     & -     &  4.3 &       & 3.08 & -42.9 \\
      & MC1    & 1.13 & 1.0     & -     &  9.6 &       & 4.23 & -30.8 \\ \hline
\end{tabular}
\end{table*}
%% Tables may also be prepared as separate files. See the accompanying
%% sample file table.tex for an example of an external table file.
%% To include an external file in your main document, use the \input
%% command. Uncomment the line below to include table.tex in this
%% sample file. (Note that you will need to comment out the \documentclass,
%% \begin{document}, and \end{document} commands from table.tex if you want
%% to include it in this document.)


%% The following command ends your manuscript. LaTeX will ignore any text

\end{document}